\documentclass[submission,copyright,creativecommons]{eptcs}
\providecommand{\event}{SOS 2007} % Name of the event you are submitting to
\usepackage{breakurl}             % Not needed if you use pdflatex only.
\usepackage{underscore}           % Only needed if you use pdflatex.
\usepackage[most]{tcolorbox}

\title{Natural Language Proof Checking in Introduction to Proof Classes -- First Experiences with Diproche}
\author{Merlin Carl
\institute{University of Flensburg\\ Flensburg, Germany}
\email{merlin.carl@uni-flensburg.de}
\and
 Hinrich Lorenzen
\institute{University of Flensburg\\ Flensburg, Germany}
\email{\quad hinrich.lorenzen@uni-flensburg.de}
\and
Michael Schmitz
\institute{University of Flensburg\\ Flensburg, Germany}
\email{\quad michael.schmitz@uni-flensburg.de}
}

\def\titlerunning{Diproche Article}
\def\authorrunning{M. Carl, H. Lorenzen \& M. Schmitz}
\begin{document}
\maketitle

\begin{abstract}
	
We present and analyze the employment of the Diproche system, a natural language proof checker, within a one-semester mathematics beginners lecture with 228 participants. The system is used to check the students' solution attempts to proving exercises in Boolean set theory and elementary number theory and to give them immediate feedback. The benefits of the employment of the system are assessed via a questionnaire at the end of the semester and via analyzing the solution attempts of a subgroup of the students. Based on our results we develop approaches for future improvements.

\end{abstract}

\section{Introduction}

In this article we present a project about engaging an automated proof checker, which reads proofs that are written in a controlled natural language, in mathematics beginners courses. The system was employed in winter term 2020/21 within a first semester mathematics lecture (and its accompanying tutorials and weekly problem series) at the University of Flensburg with 228 participants. Besides observing the benefits from the automated proof checking in general we systematically collected and analyzed data in the following way. All students were given the opportunity to participate in a questionnaire at the end of the semester and the solution attempts of 56 of the students that were submitted during the semester were analyzed. %\textcolor{red}{sollte jetzt OK sein, oder? Yup.} 
%Note that in this article we focus on analyzing the insights from lecture, tutorials, and weekly problem series. The exam that was written at the end of the course is not part of this first study.
%MC: erst einmal auskommentiert. warten wir doch ab, ob überhaupt jemand fragt...

In Section~2 we provide a brief introduction to the system. The next section describes the general structure of the corresponding beginners lecture, including some remarks from the perspective of mathematics education on the goals and the general teaching philosophy that stands behind them. Section~4 explains in detail how the system was integrated into the lecture and the corresponding tutorials. The results from the submitted students' solutions and from the questionnaire are presented and discussed in Section~5. In concluding the article, in Section~6 we share thoughts about what the benefits of the system are at the moment and what could be improved in future work.

%\section{The Diproche System}\label{sec:the_diproche_system}

\section{The Diproche System}

Diproche is an acronym for ``DIdactical PROof CHEcking''. The Diproche system is developed to check and provide automated feedback on solutions to elementary mathematical proving exercises written in a controlled fragment of German, the ``Diproche language''. A brief overview of Diproche can be found in \cite{CK20}, a more extensive discussion of how elementary number theory and axiomatic geometry are treated in Diproche has appeared in \cite{Carl20}. Here, we restrict ourselves to a short summary. 

Diproche arose out of the motivation to develop a variant of the Naproche (``NAtural language PROof CHEcking'') system due to Koepke, Schr\"oder, Cramer and others (see, e.g., \cite{Cramer}) specifically designed to proof exercises in first introductory courses to proof methods in German universities. Input texts written in a controlled natural language\footnote{In particular, the Diproche language contains several usual expressions for declaring a variable, making an assumption, announcing a proof goal, structuring the proof (e.g., by announcing a proof method) and drawing a conclusion. An account of the Diproche language can be found in \cite{Carl2021}.} designed to mirror closely the language in which (formally correct) solutions to beginner student's proving exercises are usually written are automatically formalized and then handed over to an automated theorem prover (ATP) set up in a way to accept only proof steps that can be regarded as sufficiently elementary for being used in the respective exercise. The system then provides feedback on (i) the textual input (unknown symbols/words, non-wellformed formulas, nonprocessable sentences), (ii) type errors (use of variables that were not priorly declared or use of variables in a way incompatible with their declaration, such as adding propositions or assuming a number), (iii) non-verifiable logical steps, (iv) occurence of typical error patterns, such as deducing $\neg B$ from $A\rightarrow B$ and $\neg A$ or calculating $(A+B)^2$ as $(A^2+B^2)$, (v) achievement, or otherwise, of partial proof goals (if such are declared). In certain cases, the feedback also includes counterexamples to false claims.\footnote{Counterexamples are currently provided for the fields of propositional logic and Boolean set theory. If a propositional formula is claimed to follow from a set of premises from which it does not actually follow, then an assignment of truth values is offered that makes the premises true but the conclusion false. Likewise, if a statement, e.g., of the form $x\in A$ is wrongly claimed to follow from a set of assumptions in the language of Boolean set theory, a situation is described in which this fails.} The verification by the ATP checks each supposed inference step in the text. In particular (modulo the presence of (fixable) bugs in the system), it is not possible to write false or nonsensical proof texts that are accepted by the system as valid.

In order to remain close to the common way of expressing things, the Diproche language and ATP are specifically adapted to each mathematical area in which exercises are posed. Currently, the Diproche system supports exercises in propositional logic, Boolean set theory, elementary number theory, axiomatic geometry, elementary group theory, induction and ``functions and relations''.

In our application of the system during the winter semester 2020/2021, only the modules for Boolean set theory and elementary number theory were actually used. 

For the sake of illustration, in Figure~\ref{examples} example Diproche exercises are shown that were posed during the winter term 2020/21, together with (examples of) solutions accepted by the system; further examples can be found in \cite{Carl20}.

\begin{figure}
\noindent\begin{tcolorbox}[width=.475\textwidth, nobeforeafter, title=Example Diproche Text 1]
 Es sei x eine ganze Zahl. Zeige: Wenn x gerade ist, dann ist 2-3*x gerade.

\bigskip

Beweis: Es sei x gerade. Dann gibt es eine ganze Zahl k mit x=2*k. 
Sei k eine ganze Zahl mit x=2*k. Dann ist 2-3*x=2-3*(2*k)=2*(1-3*k). 
Also ist 2-3*x gerade. qed.\footnote{Let $x$ be an integer. Prove: If $x$ is even, then $2-3x$ is even. \newline Proof: Let $x$ be even. Then there is an integer $k$ such that $x=2k$. Let $k$ be an integer with $x=2k$. Then we have $2-3x=2-3\cdot (2k)=2(1-3k)$. Hence $2-3x$ is even. qed.}
\end{tcolorbox}\hfill
\begin{tcolorbox}[width=.475\textwidth, nobeforeafter, title=Example Diproche Text 2]

 Es seien A, B, C Mengen. Wir zeigen: $((A\cap B)\times C)\subset(A\times(B\cup C))$.

\bigskip

Beweis: Sei $(x,y)\in((A\cap B)\times C)$. Dann gilt $x\in(A\cap B)$ und $y\in C$. Also gilt $x\in A$ und $x\in B$. 
Damit haben wir $y\in (B\cup C)$. Folglich ergibt sich $(x,y)\in(A\times (B\cup C))$. qed.\footnote{Let $A$, $B$, $C$ be sets. We show: $((A\cap B)\times C)\subset(A\times(B\cup C))$. \newline Proof: Let $(x,y)\in((A\cap B)\times C)$. Then it holds that $x\in(A\cap B)$ and $x\in C$. Hence we have $x\in A$ and $x\in B$. It follows that $y\in (B\cup C)$. Consequently, we have $(x,y)\in(A\times(B\cup C))$. qed.}
\end{tcolorbox}
\caption{Example Diproche exercises}\label{examples}
\end{figure}

\section{The Corresponding Beginners Lecture}

The lecture in which the Diproche system was employed has the title `Algebra I und ihre Didaktik' (`Algebra I and its Didactics') and is the only mathematics lecture that first semester students at the University of Flensburg take. All of these mathematics students are ongoing teachers for basically all forms of schools in the German education system, namely primary school, secondary school (middle and upper level), and vocational school. The word algebra in the title refers to basic algebra and not to higher algebraic theories, which means that the lecture introduces elementary fundamental principles of university mathematics, such as propositional logic, naive set theory, working with elementary algebraic expressions and equations, a few aspects of elementary number theory, and very basic aspects of group theory.

Teaching how to prove is one of the main goals of the course, and it is intended to be accomplished through these topics. The lecture aims particularly at addressing fundamental proof techniques carefully and in detail. Among the techniques that are imparted are direct proof, proof by exhaustion, proof by contradiction, proof by contraposition, and mathematical induction. Proof training is so central to the course that for instance the topic number theory is comprised only for the sake of providing good proving problems. Examples from elementary number theory are often suitable, since the mathematical techniques that are required, such as simple term manipulations, are less complicated than in other fields, so that the students can focus on general aspects of proving, namely the logical structure of the involved statements, mathematical language, and so on. Many of the chosen problems are in the vein of \cite{Chartrand}.

While in other countries transition-to-university-mathematics courses are usual, this is not the case in most German universities. Nevertheless, teaching how to prove has almost completely vanished from German school curricula. Consequently, first semester students do normally not dispose of any proving abilities. While some universities count on the students learning how to prove on the fly through the usual topics that are imparted, our institute decided to explicitly address proof techniques in the first semester classes. Therefore, some of the topics are chosen not for their mathematical content, but for their suitability for proof training. This can be understood in the vein of competency-oriented teaching of mathematics, where general abilities, such as mathematical argumentation, are central, rather than certain specific mathematical contents. Of course, care is taken to ensure that basic mathematical notions, which are needed for the further course of studies, are imparted. Moreover, care is taken to choose the right level of difficulty, namely to consider topics and problems that are neither trivial nor extremely hard. In this regard it is worth reminding that all mathematics students at the University of Flensburg are ongoing teachers. It is sometimes said that the topics that are taught at university are not relevant to school mathematics. Of course, we strongly disagree. Much of university mathematics is immediately relevant to contents that are taught in school, just think of the importance of propositional logic and naive set theory for working with equations. But there is a grain of truth, that can be formulated in a positive manner: For some of the topics of  university mathematics, it is not so important which of these are chosen to educate ongoing teachers. For instance, it does not matter if a teacher student specializes in advanced calculus or if they focus on higher algebraic theories or axiomatic geometry. It is only important that a teacher student is introduced to a serious mathematical theory at some point in time to experience how mathematical theories evolve. From these considerations, we gain some degree of freedom for educating mathematics teachers. This freedom is used to focus on imparting how to think mathematically through seriously working on elementary but not trivial topics.

Within teaching mathematical argumentation, the use of mathematical language plays a dominant role and can be seen as a major hurdle for the students (cf, e.g., \cite{Mejia-Ramos}). Mathematical language is extremely precise and condensed (see \cite{Schroeder}) and makes use of formal expressions and symbols, which do not play a role in non-mathematical language. Both for reading and writing mathematical texts it is necessary to translate from textual to symbolic expressions and vice versa. Moreover, mathematical phrases and the grammar inherent to mathematics must be used in a correct way. This is not only necessary for being able to please the readers of one's owns mathematical texts, but to achieve the ability to work successfully in mathematics. This is the case, because only then the benefit of a condensed and precise language and symbolism, namely to express facts in a concise manner that disburdens the mind, is accessible. To this end, repeated and purposeful training in reading and writing mathematical texts plays a major role in the here described first semester course. It seems that Diproche can be a helpful companion to achieve its goals.

\section{Integration of the Diproche System to Lecture and Tutorials}

As mentioned above, the Diproche System was tested within the first-semester course `Algebra I and its Didactics' given by Prof. Hinrich Lorenzen in winter semester 2020/21 at the University of Flensburg. The course consists of a four-hour \textit{lecture} (two sessions a week, held by the professor), weekly \textit{problem series}, weekly two-hour \textit{tutorials} (the tutors being scientific associates), and weekly two-hour \textit{preparatory tutorials} (accompanied by the same tutors). A problem series contains about 4-6 problems, which the students are asked to solve in teams of two persons within a week and then submit their handwritten solutions. The students' submissions are marked up by student assistants from higher semesters. Within the preparatory tutorials, which take place in the same week a problem series is disclosed, the students work on the problems in small groups while a tutor is present to answer questions or to give small hints. Within the tutorials, which take place a week later, the solutions to the problems are discussed. It is desired that students come to the blackboard and present their solutions, but sometimes it is necessary that the tutor presents a solution. This usual procedure was adapted to integrate the Diproche system in fall term 2020/21 as will be described below. It should be noted that due to the Corona pandemic all above mentioned teaching events had to be conducted online via the tool Cisco Webex Meetings.

For the sake of scientific evaluation of the benefits of the Diproche System it was first thought to have only a subgroup of the students to be introduced to the system, while the rest of the students serve as a control group. Since `Algebra I and its Didactics' is a regular compulsory course, in which everybody should have equal chances, this idea was discarded for ethic reasons. It was decided to rather introduce the Diproche system to all students in the following way. First, several passages in the lecture that were considered suitable to demonstrate the system had to be identified and the demonstrations were then given by the professor within the lecture. In addition, a workshop was conducted to introduce the Diproche system to the tutors. The problem series were supplemented by so called `Diproche problems', that were intended to be solved by the students by using the Diproche system.  In order not to place an excessive burden on the student assistants it was decided not to mark the student solutions to the Diproche problems by hand. The Diproche problems were developed by the authors and some of them were tested by the tutors before disclosing them on a problem series. These tests proved to be beneficial, as thereby minor problems and uncertainties could be resolved in advance. This approach could be established for all Diproche problems in future runs, when hopefully no Corona restrictions impede the exchange between colleagues.

%It was intended to have the Diproche problems (that were developed by the authors) tested by the tutors one week before disclosing them on a problem series. Since the exchange between colleagues was impeded by the impact of the Corona pandemic this idea was not feasible, so the Diproche problems had to be disclosed without the desired lead time, which sometimes caused minor problems and uncertainty on the side of the tutors and students. Moreover, problems with setting up the web server resulted in the system not being able to be used by the students until the fifth week of the semester.

Technical problems with setting up the web server resulted in the system not being able to be used by the students until the fifth week of the semester. Therefore and for the sake of introducing the system gently, the Diproche problems served mainly as repetitions of techniques that were already imparted, sometimes by using slightly different notations and formulations than the one Diproche expects. It could be beneficial (and probably easliy feasible) to avoid both notation differences, such as $\sim$ instead of $\neg$, and differences in formulations in future runs to introduce the Diproche syntax to the students as a first approach to mathematical terminology. This would be a possible way to disguise the standardization that comes with the Diproche CNL. Note that this does not mean that no alternative formulations are usable, since the Diproche CNL possesses a variety of alternative formulations for most situations.

% are caused by the fact that the Diproche GUI is somewhat limited until now, the differences in formulations could have been avoided, if the exchange between colleagues would not have been impeded by the pandemic. These circumstances undermined the idea to introduce the Diproche syntax to the students as a first approach to mathematical terminology. This is unfortunate as it would have a been a possible way to disguise the standardization that comes with the Diproche CNL. Summarizing, both the tutors and the students have not been introduced to the Diproche system to the desired extent. Nevertheless, several good suggestions to improve the system came from this test run-through, such as new formulations or inferences that should be accepted by the system.

For the sake of the students the Diproche problems on the weekly problem series were given by a link to their implementation within the system rather than by writing out the problem text. In addition to the explicitly marked Diproche problems some of the usual proof problems (that seemed particularly suitable) were also provided with a link and a remark that they could be done by using Diproche. The students were encouraged to call in if they had any problems with the system. Because there were few requests, and these were mostly related to the Diproche feedback to particular solution attempts, we assume that the system was accessible to the students at any time and that there were no instabilities of the web server or so. In winter semester 2020/21 there were 228 participants, so that we had 114 submissions by teams of two students a week, of which 28 were chosen for evaluation. For the sake of illustration, we now consider and briefly discuss one of the Diproche problems that was employed within the course.

\begin{center}
\noindent\begin{tcolorbox}[width=.9\textwidth, nobeforeafter, title=Diproche problem from series 8]
 Es sei x eine ganze Zahl. Zeige: Wenn x\^{}2+2 ungerade ist, dann ist x ungerade.

\bigskip

Beweis. Angenommen, x ist nicht ungerade. Dann ist x gerade. Also existiert eine ganze Zahl k mit x=2*k. Sei k eine ganze Zahl mit x=2*k. Dann folgt x\^{}2+2=(2*k)\^{}2+2=4*k\^{}2+2=2*(2*k\^{}2+1). Also ist x\^{}2+2 gerade. Also ist x\^{}2+2 nicht ungerade. qed.\footnote{Let $x$ be an integer. Prove: If $x^{2}+2$ is odd, then $x$ is odd. \newline 
Proof: Assume that $x$ is not odd. Then $x$ is even. Hence there exists an integer $k$ such that $x=2k$. Pick an integer $k$ such that $x=2k$. Then we have $x^2+2=(2*k)^+2=4*k^+2=2*(2*k^2+1)$. Hence $x^{2}+2$ is even. It follows that $x^{2}+2$ is not odd. qed.}
\end{tcolorbox}
%\textcolor{red}{Darstellung ab $x^2$ komisch. Sollen wir eine Übersetzung ergänzen?}
\end{center}

This problem is from the part of the lecture where proof by contraposition is imparted through very elementary number theory. It can be seen that the problem urges the student to choose the desired proof method rather than asking him or her to do so. Many students would first start by assuming that $x^2+2$ is of the form $2k+1$ and then run into problems. It is intended that the students do so and then analyze where the problem lies and think about how another proof method could avoid the obstacle. As mentioned above, the idea is to choose problems with low complexity, so that the students can focus on the structure of their proof and on the way in which it is formulated. Note that the form of a proof by contraposition was adhered to very strictly in this proof text, so that, e.g., rather than starting with the assumption that $x$ is even, we assume that $x$ is not odd and then deduce that $x$ is even.
%The system allows a certain degree of freedom. For instance, instead of first assuming that $x$ is not uneven and concluding in the next sentence that is has to be even in order to match precisely the structure of a proof by contraposition, one could have assumed right from the beginning that $x$ is even. [stimmt nicht]

\section{Results}

In this section, we summarize some of the quantitative results obtained from a survey of the usage of Diproche, concerning (i) the amount of Diproche exercises to which students handed in solutions and the success rates for these exercises and (ii) the results of a questionnaire in which the system was evaluated from the student's point of view. 

 \subsection{Frequency and Quality of Submitted Solutions to Diproche Exercises}

The exercises were worked on and handed in teams of two students. For the sake of our study, $28$ of these teams were considered over the semester. 

Concerning proof exercises on elementary number theory, $14$ Diproche exercise problems were posed during the semester; for these, a total of $184$ solution attempts were handed in, which amounts to an average of $13.14$ attempted solutions per problem. Thus, the average Diproche exercise on number theory was attempted by about half of the teams. Of these $184$ attempted solutions, $104$, or about $57\%$, were reported by the system as correct, while an additional $29$, or about $16\%$ were flagged as having minor mistakes but might have been considered correct by a human corrector. A frequent reason for this was that students forgot to explicitly state the goal of the exercise as the last sentence of their solution and instead ended with something from which the proof goal easily follows, which resulted in their solution being flagged as not reaching the proof goal. Thus, each exercise received on average about $7.4$ solutions that were reported as correct by the system, and an additional $2$ that were ``almost correct''. With respect to the total number of students, this means that the average exercise was solved correctly by about $27\%$ of all students.\footnote{Note that this number takes into account those students that did not hand in a solution.}

With respect to proof exercises on Boolean set theory, there were $10$ exercises that received a total of $86$ solutions; thus, the average problem was attempted by $8.6$ of the teams. Of these attempted solutions, $44$, or about $51\%$, were reported by the system as correct, while another $14$, or about $16\%$, were reported as having minor mistakes but might have been considered correct by a human corrector. Thus, the average proof exercise for Boolean set theory received about $4.4$ correct and an additional $1.4$ ``almost correct'' solutions. With respect to the total number of students (including those who did not hand in solutions to these exercises), this means that the average exercises was solved correctly by about $16\%$ of the students.

An encouraging observation is that few solutions were merely ``almost correct'' in the sense that the system reported a mistake while human correctors might have considered the solution correct. This shows that the standardization of natural language and inference rules introduced by the system does not impede the use of the system too strongly: The system is not perfectly natural, but ``sufficiently natural''. This is also confirmed by the fact that a reasonable portion of the students handed in correct solutions without having received a systematic introduction into the use of the system, merely by means of having been shown several correct examples texts. A particularly encouraging observation in this respect is the outcome of the exercises on divisibility on exercise sheet $10$; these exercises were posed although no example proof about divisibility, and no explanation of the respective syntax or language, had been priorly introduced; still, the number of solutions for two of these exercises\footnote{These exercises were to show that $8$ divides $(2n-1)^2-1$ for all integers $n$ and that $4$ divides $3n^2$ for even $n$} written in correct and processable Diproche language did not deviate from the respective number for the  other problems. This further confirms that the Diproche language is ``natural'' for its intended range of applications.

On the other hand, the relatively high number of students who omitted Diproche exercises -- the average Diproche exercise was attempted by about half the students -- is somewhat unfortunate. One reason may be that the students did try to solve these problems, but did not submit their work after the system had reported it as incorrect. 
%Unerfreulich: hohe zahl von nichtbearbeitungen. m\"oglich, dass hier versuche unternommen, aber die als nicht korrekt gemeldeten l\"osunge nicht abgegeben wurden. 

Similarly unfortunate is the frequency of submitted solutions with rather basic mathematical mistakes in spite of these being reported by the system; for example, several solutions contained elementary term manipulation mistakes such as $(2z)^{3}=2z^{3}$, which the system flagged as incorrect. This indicates that students had difficulties to use the system's output for improving their solutions, which is further confirmed in the survey below.

\subsection{Student's Evaluation\protect\footnote{This section is an abbreviated English version of the first author's thesis \cite{Carl2021}.}}

The evaluation questionnare consisted of four free-text questions and $28$ multiple-choice questions, covering the topics of (i) user-friendliness and usability, 
(ii) nature and extent of use, (iii) effect on the understanding of proofs, (iv) effect on the competence of constructing and writing proofs, (v) effect on motivation and self-confidence, (vi) quality of integration into the teaching process. Some of the questions asked were inspired by the discussion of the evaluation of Lurch in Carter and Monks \cite{CM}. Each of these questions was to be answered by a natural number ranging from (1) (complete disagreement) to (6) (complete agreement). The questionnaire was answered by $127$, or about $56\%$, of the $228$ participants of the lecture. Here, we summarize some of the findings. 

The free-text questions included the questions (i) what the students particularly liked and (ii) what they particularly disliked about the system. 

Of the $49$ answers we received to question (i), $43$ concerned one or several of the following points; the frequency with respect to the $49$ answers we recieved (rounded to the nearest integer) is given in the brackets after the answer.

\begin{enumerate}
\item Diproche provides an immediate feedback. ($58\%$)
\item Diproche shows faulty passage in proof texts. ($16\%$)
\item Diproche allows one to work interactively on a proof, using the corrections. ($12\%$)
\item Diproche makes it possible to correct solutions without the help of others. ($16\%$)
\item Diproche leads to increased trust with respect to the correctness of one's proof texts. ($5\%$)
\item Diproche has an effect on learning the language of mathematics. ($7\%$)
\item Diproche has an effect on learning strategies and structures of mathematical proofs. ($9\%$) 
\item Diproche has a positive effect on the motivation to work on proofs. ($5\%$) 
\end{enumerate}

The remaining six answers either concerned the improvements to the system made during the semester or general (positive) feedback about the usefulness of the system. 

For question (ii), we received $57$ answers, $54$ of which fell under one or several of the following points, whose frequency among the answers received is given in brackets after the answer: 

\begin{enumerate}
\item It is difficult to find mistakes in the proof text on the basis of the error messages of the system. ($67\%$)
\item The corrections of the system are too ``meticulous'', for example, texts cannot be processed when they contain typos or punctuation mistakes. ($28\%$)
\item Diproche has a negative influence on the motivation to work on proofs. ($12\%$)
\item The linguistic standardization increases the difficulty in writing proof texts. ($12\%$) 
\item The logical standardization increases the difficulty in writing proof texts. ($7\%$) 
\item It is too time-consuming to write a proof text accepted by Diproche. ($28\%$) 
\item The correctness norm for proof texts used by Diproche deviates from the one used in the lectures and the other exercises. ($2\%$)
\item The explanations about the system in the lecture and the exercises were insufficient. ($12\%$)  
\item The linguistic and syntactical standardization of Diproche deviates from that used in the lectures and the exercise classes. ($9\%$) 
\item Diproche does not offer sufficient support in searching for solutions. ($2\%$) 
\end{enumerate}

Concerning the user-friendliness of the system, the following questions were asked:

\begin{enumerate}
\item It is easy for me to enter a text into the Diproche system. %(1)
\item Diproche-proofs resemble proofs that I know from lectures and exercise classes. %(2)
\item If the system reports a mistake, I can often recognize what's wrong in my proof text. %(10)
\item The feedback that Diproche provides helps me to improve my proof texts. %(17) 
\end{enumerate}

The number of answers and their distribution and the average value can be found in table \ref{tab1}. %Die Anzahl der Antworten, die Verteilung der Antworten sowie der Durchschnittswert ergeben sich aus folgender Tabelle:

\begin{table}[htbp]
\caption{Student's assessment of user-friendliness}\label{tab1}
\resizebox{\columnwidth}{!}{
\begin{tabular}{|l|c|c|c|c|c|c|c|c|}
Question & Answers & 1 & 2 & 3 & 4 & 5 & 6 &  Average \\
\hline
1 & 125 &  13 (10\%) & 40 (32\%) & 26 (21\%) & 29 (23\%) & 12 (10\%) & 5 (4\%) & 3.0 \\
\hline
2 &125 & 3 (2\%)      & 13 (10\%)  & 31 (25\%) & 35 (28\%) & 35 (28\%) & 8 (6\%) & 3.9\\
\hline
3 &124&  25 (20\%) &   47 (38\%) & 28 (23\%) &15 (12\%) &   8  (6\%)   & 1 (1\%) & 2.5\\
\hline
4 &124&  15 (12\%) &  36 (29\%) &  33 (27\%)& 31 (25\%) &   7  (6\%)  & 2 (2\%) & 2.9\\
\hline
\end{tabular}
}
\end{table}

Considering a response of $3$ or less as ``negative in tendency'' and one of $4$ or higher as ``affirmative in tendency'', we see that a majority of respondents ($63\%$) indicate that typing diproche texts is not easy for them. This could be due to difficulties with the formulation of proof texts on the one hand, but also due to additional difficulties raised by the normalizations imposed by the system. That some aspect of the standardization (such as the linguistic, logical, structural or syntactic) made the writing of Diproche texts more difficult was mentioned by as many as 26\% of the free-text answers, although only 12\% explicitly referred to the linguistic standardization. On the other hand, a clear majority (62\%) perceived Diproche proof texts as similar to the proof texts from lectures and exercises; this is also in line with the free text answers, where only 9\% complained about a deviation of the linguistic standardization of Diproche from that of lectures or exercises. The linguistic standardization thus seems to represent an additional obstacle to the writing of proof texts, but not an insurmountable one. This is consistent with the experience from the submitted exercises that a clear majority of the submissions to the first Diproche proof exercises on elementary number theory were formally correct in the input language of Diproche.

Significantly greater problems show up in the comments on the system's feedbacks: Only 19\% reported being able to identify errors in their proof text when given Diproche error messages, and only about a third (33\%) of students were able to use Diproche feedback to improve a proof text. This is also clearly reflected in the free-text responses: the difficulty of finding the errors in the text using the feedback was mentioned in 67\% of the answers given; 33\% of the answers to the question about suggestions for improvement concerned the error messages. This is further in line with the results of the data collection of the submissions, where transformation errors such as $(2x)^3=2x^3$ were frequently reported by the system, but not corrected by the students.

Concerning the extent to which the system was used, the questionnaire contained the following questions, where the second was a free-text question:

\begin{enumerate}
\item Due to Diproche, I  spent more time on proving exercises than I would have done for the same problems without using the system.
%\item Ich habe mich auch \"uber die gestellten \"Ubungsaufgaben hinaus mit Aufgaben im Rahmen von Diproche besch\"aftigt. (16)
%\item Auch Beweise, die ich f\"ur richtig halte, gebe ich manchmal in Diproche ein, weil es mir Freude macht. (24)
\item Please estimate roughly how much time you spent with the Diproche system per week on average.  
\end{enumerate}

Concerning question $1$, we received the replies in table \ref{tab2}.

\begin{table}[htbp]
\caption{Student's assessment of extent and way of use}\label{tab2}
\resizebox{\columnwidth}{!}{
\begin{tabular}{|l|c|c|c|c|c|c|c|c|}
Question & Answers & 1 & 2 & 3 & 4 & 5 & 6 & Average \\
\hline
1 &125& 12 (10\%) & 9 (7\%) & 13 (10\%) & 17 (14\%) & 34 (27\%) & 40 (32\%) & 4.4\\
\hline
%2 & 124 & 57 (46\%) & 33 (27\%) & 20 (16\%) & 8 (6\%) & 4 (3\%) & 2 (2\%) & 2.0\\
%\hline
%3 & 124 & 76 (61\%) & 24 (19\%) & 12 (10\%) & 7 (6\%) & 4 (3\%) & 1 (1\%) & 1.7\\
%\hline
\end{tabular}
}
\end{table}

Concerning the average time per week spent on the exercises, we received $86$ usable replies, which yielded the distribution in table \ref{tab3}.

\begin{table}[htbp]
\begin{center}
\caption{Student's report on average time per week spent on the Diproche exercises}\label{tab3}
%\resizebox{\columnwidth}{!}{
\begin{tabular}{|c|c|c|c|c|c|c|}
Hours & $0-1$               & 1-2                     & 2-3                       & 3-4                & 4-5                     & ${>}5$ \\
\hline
Number     &  $19$ ($22\%$) &$31$ ($36\%$)   & $26$  (30\%)      & $5$  (6\%)    &  $4$  ($5\%$)      & $1$ ($1\%$)\\
\hline
\end{tabular}
%}
\end{center}
\end{table}

Thus, the average time per week spent on Diproche exercises was between $2$ and $2.4$ hours. This means that the students spent between about $20$ minutes on each of the Diproche exercises on average.

Concerning the effect on understanding proofs, the questionnaire contained the following question, the replies to which are summarized in table \ref{tab4}.

%wirkung auf verständnis, fr. 3 (2. 276)

%Zur Wirkung auf das Verst\"andnis von Beweisen wurden folgende Fragen gestellt (die Nummer der Frage im Evaluationsbogen ist jeweils in Klammern hinter der Frage angegeben): 

\begin{enumerate}
%\item Ich verstehe den Unterschied zwischen Logik- und Typenfehlern. (15)
%\item Die Arbeit mit Diproche hilft mir dabei, besser zu verstehen, wie Beweise funktionieren.\footnote{Vgl. f\"ur diese und die folgende Frage Carter und Monks  \cite{CM}, S. 8 in der Evaluation von Lurch, wo in einer Freitextantwort angemerkt wurde, Lurch habe dabei geholfen, zu verstehen, was ein Beweis ist.} (20) 
\item Working with Diproche helps me to improve my understanding of proofs that appear in lectures and exercises.  
\end{enumerate}

%(1) erfasst, ob Studierende zwischen einer Fehlverwendung von Variablen und Schlussfehlern unterscheiden k\"onnen. 

\begin{table}[htbp]
\caption{Student's assessment on the effect of the system on proof understanding}\label{tab4}
\resizebox{\columnwidth}{!}{
\begin{tabular}{|l|c|c|c|c|c|c|c|c|}
Question & Answers & 1 & 2 & 3 & 4 & 5 & 6 & Average \\
\hline
%1 &125&7 (6\%) & 13 (10\%) & 34 (27\%) & 26 (21\%) & 24 (19\%) & 21 (17\%) & 3.9 \\
%\hline
%2 &125 & 13 (10\%) & 31 (25\%) & 31 (25\%) &39 (31\%) & 10 (8\%) & 1 (1\%) & 3.0 \\
%\hline
3 &124 & 18 (15\%) & 40 (32\%) & 33 (27\%) & 26 (21\%) & 6 (5\%) & 1 (1\%) & 2.7 \\
\hline
\end{tabular}
}
\end{table}

Thus, only a bit more than a fourth ($27\%$) of the students reported a positive effect of the system on proof understanding.

About the effect on proving competences, i.e., finding, checking, correcting and presenting proofs, the following questions were among those asked in the questionnaire, the answers to which are summarized in table \ref{tab5}.

%wirkung auf beweiskompetenz: fr. 4-7

%Zur Wirkung auf die eigenen Beweiskompetenzen  -- also eigene Beweise zu finden, kritisch zu betrachten, ggf. zu korrigieren und darzustellen -- wurden folgende Fragen gestellt (die Nummer der Frage im Evaluationsbogen ist jeweils in Klammern hinter der Frage angegeben): 

\begin{enumerate}
%\item Diproche hilft mir, Probleme in meinen Beweisans\"atzen zu erkennen. (3)
%\item Durch Diproche habe ich etwas dar\"uber gelernt, wie man einen Beweistext hinsichtlich seiner Korrektheit beurteilt. (4) 
%\item Diproche hat mir dabei geholfen, mich daran zu gew\"ohnen, Variablen vor ihrer Verwendung (mit Formulierungen wie ``Sei x eine ganze Zahl." o.\"a.) einzuf\"uhren. (5)
\item Diproche has helped me to recognize and avoid false inferences steps. %Diproche hat mir geholfen, falsche Schlussweisen zu erkennen und zu vermeiden. (6)
\item Working with Diproche helped me to learn the mathematical vernacular. %Die Arbeit mit Diproche hilft mir dabei, die mathematische Fachsprache zu lernen. (22)
\item Working with Diproche helped me to learn how to structure proof texts. %Die Arbeit mit Diproche hilft mir, zu lernen, wie man Beweistexte strukturieren kann. (23)
%\item I learned a lot by working with Diproche. %Ich habe durch die Verwendung von Diproche viel gelernt. (25)
%\item Um Diproche-Aufgaben zu l\"osen, muss man die zu erwerbenden Kompetenzen bereits besitzen; daher ist Diproche nutzlos. (26)

\end{enumerate}

\begin{table}[htbp]
\caption{Student's assessment of the effect of Diproche on proving competences}\label{tab5}
\resizebox{\columnwidth}{!}{
\begin{tabular}{|l|c|c|c|c|c|c|c|c|}
Question & Answers & 1 & 2 & 3 & 4 & 5 & 6 & Average \\
%\hline
%1 & 125 & 27 (22\%) & 33 (26\%) & 30 (24\%) & 24 (19\%) & 10 (8\%) & 1 (1\%) & 2.7\\
%\hline
%2 & 124 & 14 (11\%) & 21 (17\%) & 36 (29\%) & 37 (30\%) & 14 (11\%) & 2 (2\%) & 3.2\\
%\hline
%3 & 123 & 13 (11\%) & 13 (11\%) & 18 (15\%) & 24 (20\%) & 39 (32\%) & 16 (13\%) & 3.9\\
\hline
1 &125 & 17 (14\%) & 36 (29\%) & 42 (34\%) & 21 (17\%) & 8 (6\%) & 1 (1\%) & 2.8 \\
\hline
2 &125 & 8 (6\%) & 14 (11\%) & 27 (22\%) & 47 (38\%) & 25 (20\%) & 4 (3\%) & 3.6\\
\hline
3 &125 & 9 (7\%) & 11 (9\%) & 27 (22\%) & 40 (32\%) & 30 (24\%) & 8 (6\%) & 3.8\\
\hline
%4 &125 &18 (14\%) & 33 (26\%) & 35 (28\%) & 30 (24\%) & 9 (7\%) & 0 (0\%) & 2.8\\
%\hline
%8 &124 & 9 (7\%) & 31 (25\%) & 36 (29\%) & 28 (23\%) & 15 (12\%) & 5 (4\%) & 3.2 \\
%\hline
\end{tabular}
}
\end{table}

%Die Frage, ob das System dabei hilft, sich an die Einf\"uhrung von Variablen zu gew\"ohnen, wurde von 65\% der Befragten \"uberwiegend bejaht. 

%WEITER HIER!

As can be seen from table \ref{tab5}, a positive effect of the system on the linguistic and linguistic-related aspects of proving (use of vernacular (2), structuring expositions (3)) is reported by a majority of the students (61\% and 62\% gave answers in the range $4-6$, respectively). 
For the logical aspects, this was only the case for a minority (24\%). 

Finally, concerning the effect of Diproche on self-confidence the motivation to deal with proofs, our questionnaire contained the following question: 
%motivation: fr. 3 (s. 279)

%Zur Wirkung auf Motivation und Selbstbewu{\ss}tsein wurden folgende Fragen gestellt (die Nummer der Frage im Evaluationsbogen ist jeweils in Klammern hinter der Frage angegeben):\footnote{Die folgenden Fragen sind u.a. angeregt durch die in Carter und Monks \cite{CM} auf S. 8 zitierten Freitextantworten von Studierenden aus der Evaluation von Lurch zu diesem Aspekt.} 

\begin{enumerate}
%\item Durch den Umgang mit Diproche bin ich im Umgang mit Beweisen sicherer und selbstbewusster geworden. (8)
%\item Diproche hat mein Vertrauen in meine eigenen Beweistexte erh\"oht. (9) 
\item I feel encouraged when the system reports a solution as correct. %Es ermutigt mich, wenn das System eine L\"osung als korrekt meldet. (11)
%\item Es ermutigt mich, wenn das System eine verbesserte L\"osung als besser meldet als den vorherigen Versuch. (12)
\item I am discouraged when the system reports a mistake. %Fehlermeldungen durch das Systen entmutigen mich. (13)
%\item Die Arbeit mit Diproche macht mir Spa{\ss}. (14)
\end{enumerate}

%Die Anzahl der Antworten, die Verteilung der Antworten sowie der Durchschnittswert ergeben sich aus folgender Tabelle:

\begin{table}[htbp]
\caption{Student's assessment of the effect on motivation and self-confidence}\label{tab6}
\resizebox{\columnwidth}{!}{
\begin{tabular}{|l|c|c|c|c|c|c|c|c|}
Question & Answers & 1 & 2 & 3 & 4 & 5 & 6 & Average \\
%\hline
%1 &124 & 26 (21\%) & 43 (35\%) & 20 (16\%) & 28 (23\%) & 5 (4\%) & 2 (2\%) & 2.6\\
%\hline
%2 &125 & 27 (22\%) & 41 (33\%) & 25 (20\%) & 26 (21\%) &5 (4\%) &1 (1\%) & 2.6 \\
 \hline
1 &125 & 3 (2\%) & 6 (5\%) & 6 (5\%) & 20 (16\%) & 42 (34\%) & 48 (38\%) & 4.9 \\
\hline
%4 &125 & 5 (4\%) & 10 (8\%) & 12 (10\%) & 44 (35\%) & 34 (27\%) & 20 (16\%) & 4.2\\
%\hline
2 &124 & 16 (13\%) & 30 (24\%) & 28 (23\%) & 17 (14\%) & 27 (22\%) & 6 (5\%) & 3.2 \\
\hline
%6 & 125 & 33 (26\%) & 42 (34\%) & 22 (18\%) & 20 (16\%) & 8 (6\%) & 0 (0\%) & 2.4 \\
%\hline
\end{tabular}
}
\end{table}

Thus, a vast majority (88\%) reported to be encouraged when the system reports a solution as correct, while a tendency towards discouragement by error message was reported by less than half of this, namely 41\%. The averages (4.9 vs. 3.2) further confirm that the positive effect of success messages clearly surpasses the negative effect of error messages. 
Given the possibility to further correct a solution reported as incorrect, this is plausible. The numbers also indicate that the majority of the students took the system's message seriously as a report about the correctness of their attempted solution. 

%Etwa ein Drittel (35\%) der Befragten gab demnach eine \"uberwiegend bejahende ($4$-$6$) Antwort auf die Frage, ob der Umgang mit dem Diproche-System sie im Umgang mit Beweisen sicherer und selbstbewusster gemacht hat; ein erh\"ohtes Vertrauen in die eigenen Beweistexte hatte die Nutzung von Diproche bei etwa einem Viertel (26\%) zur Folge. 
%Eine deutliche Mehrheit von 88\% empfand durch das System als korrekt gemeldete L\"osungen als ermutigend, 38\% antworteten hierauf sogar mit ``stimme voll zu'' ($6$); ebenso ist die Meldung, dass eine L\"osung sich verbessert habe, f\"ur 78\% tendenziell ermutigend ($4$-$6$). Deutlich weniger, aber immerhin 41\%, werden durch Fehlermeldungen des Systems tendenziell entmutigt ($4$-$6$). Ein knappes Viertel (24\%) der Befragten gab an, dass die Arbeit mit Diproche ihnen tendenziell Spa{\ss} mache ($4$-$6$); etwas mehr (26\%) beantworteten diese Frage jedoch mit ``trifft \"uberhaupt nicht zu'', wohingegen die Option ``trifft voll zu'' von keiner und keinem der Befragten angegeben wurde. 
%
%Insgesamt zeigt sich damit, dass positive Meldungen durch das System bei einem gr\"o{\ss}eren Anteil der Befragten einen ermutigenden Effekt haben als durch Fehlermeldungen entmutigt werden; eine Steigerung des Vertrauens in die eigenen Beweistexte und ein selbstbewussterer Umgang mit Beweisen zeigte sich nur bei einer Minderheit von etwa einem Viertel bzw. einem Drittel. Freude an der Arbeit mit dem System hat nur eine Minderheit der Befragten, wogegen eine klare Mehrheit dies deutlich verneint ($1$-$2$). 

\section{Conclusion and Further Work}

The results discussed in the previous section show that the Diproche system can be used in a beneficial way by students, even with only a small amount of explicit instructions regarding the system. It seems that the Diproche language is sufficiently natural, so that the system can be used intuitively. The positive motivational effect from getting an affirmative feedback from the system outweighs the discouraging effect from getting a negative feedback. Furthermore, the students appreciate the fact that the system gives instant feedback on their solution attempts. These results encourage the further development of Diproche and its future employment in university teaching. In order to improve the system and the way it is employed we now want to analyze the critical aspects from the above presented study results.

We structure the analysis by classifying the critical insights in three categories, namely (I) Problems with understanding the Diproche feedback,  (II) Problems with distinguishing between mathematical errors and special characteristics of the Diproche system, and (III) Problems with the standardization of the Diproche language.

\bigskip

(I) The above presented results show that the students have problems to understand how the feedback that Diproche gives them can be helpful to improve their solution attempts. One aspect in this regard seems to be that the students often cannot locate their mistakes in the proof text. Thus, it could be helpful to teach them standard debugging strategies, such as working stepwise, checking from time to time, splitting long formulas into smaller parts, checking mathematical and syntactical aspects separately, and so on. Moreover it seems to be promising to have the students working with the system while a tutor is in the room, who can show them how to improve a solution attempt by locating the problems via the Diproche feedback. This measure should be easily feasible when there are no Corona restrictions any more. In addition, one could improve the visual appearance of the interface and amend the feedback by further explanations.

\bigskip

(II) Some responses give the impression that the students do not trust the system, i.e., they sometimes ascribe an error message to peculiarities of the system, although they actually made a mathematical mistake. To counteract this mistrust, it would be useful, on the one hand, to practice understanding the error messages and to make the system's messages even clearer, as explained above. On the other hand, it could be a good measure to have the students' solutions be marked up by a person in addition. In this way it could be demonstrated how similar the human assessments and the computer assessments of the solutions are.\footnote{The positive effects of additionally hand-marking solutions checked by an automated proof tutoring system has been emphasized in the evaluation of Lurch, see \cite{CM13}, p. 9.}

\bigskip

(III) We have seen that the (small) deviation between the standardized Diproche language and the mathematical language used in the lecture sometimes unsettles the students. This becomes apparent in two aspects, namely in the usage of different symbols and in slightly different formulations. The present Diproche interface uses $\varepsilon$ instead of $\in$, $\sim$ instead of $\neg$, $\&$ instead of $\wedge$, and v instead of $\vee$. While for the expert these differences seem innocuous, for the beginner they can be confusing, so it should be intended to eliminate these makeshift symbols in future versions of the interface. Regarding the second aspect, it would be helpful to better dovetail the lecture with the Diproche CNL. As mentioned above, it would be a chance to impart the Diproche language as the first approach to mathematical language for the students to disguise the standardization of the Diproche language. Of course, this requires the Diproche language to offer a sufficient variety of formulations for those mathematical phrases that are relevant to the contents of the lecture (which is essentially achieved already).\footnote{It is, of course, always possible to further extend a CNL to include more formulations. However, experience shows that a larger language does not necessarily yield an improved ``naturalness'' of the system in the sense that it is easy and intuitive to use. Adding a new (kind of) expression often leads to the natural expectation that several other expressions should also be available, which in turn lead to further expectations etc., so that the distinction between processable and nonprocessable expressions becomes less intuitive. Setting up an appropriate CNL for a system like Diproche thus seems to be more an issue of a delicate balancing rather than a question of ``the more the better''.} Provided this, by carefully concerting the course and the Diproche language a more beneficial employment of the system could be achieved.

\bigskip

Another measure to improve the benefit of the employment of Diproche that is related to all three above discussed categories would be the implementation of further exercise formats. One could present a proof text to the students and ask them to predict the Diproche feedback. As another format one could present both a proof text and a feedback and ask them to explain the feedback. Note that this is not identical to letting the students explain the feedback to their own solution attempts, since they would tend to defend their own solution while for a solution of somebody else they could plausibly rather %better 
take a neutral perspective. Another good exercise format would be to present incomplete or faulty solution attempts to the students and ask them to correct or complement. Exercise formats like these are well-established in computer programming courses, which makes it very plausible that they would also be beneficial when learning %to work 
mathematics with Diproche.

To summarize, automated proof checking, even when applied to `natural' texts, establishes new (syntactical and logical) standards of correctness, that need not necessarily agree with those established in lectures and exercises. Such deviations are a potential cause of confusion and therefore need to be noted and dealt with carefully. This applies both to technical aspects and aspects of teaching. We have experienced the advantages and prospects as well as the subtleties of using natural language proof checking in beginners' `introduction to proof' classes, but on the whole we regard our experiences as encouraging. 

\bigskip

\bibliographystyle{eptcs}

\end{document}